\documentclass[aps,prl,preprintnumbers,showpacs,twocolumn,nofootinbib]{revtex4} 
\usepackage{graphicx}
\usepackage{dcolumn}
\usepackage{bm}
\def\lsim{\raisebox{-.4ex}{$\stackrel{<}{\scriptstyle \sim}$\,}}

\usepackage{latexsym}
\begin{document}
\newcommand{\Od}{{\cal O}}
\newcommand{\A}{{\mathcal{A}}}
\newcommand{\dA}{\delta{\mathcal{A}}}

\input epsf \renewcommand{\topfraction}{0.8}

\title{A cosmic vector for dark energy}
\author{Jose Beltr\'an Jim\'enez and Antonio L. Maroto}
\affiliation{Departamento de  F\'{\i}sica Te\'orica,
 Universidad Complutense de
  Madrid, 28040 Madrid, Spain}

\date{\today}

\begin{abstract}
 In this work we show that the presence of a  vector field
on cosmological scales could explain the present phase of 
accelerated expansion of the universe. 
The proposed  theory
contains no dimensional parameters nor potential
terms and does not 
require unnatural initial conditions in the early universe, thus avoiding 
the so called cosmic coincidence problem. In addition, 
it  fits the data from high-redshift supernovae with excellent precision,  
 making definite predictions for cosmological parameters. 
Upcoming observations will be able to clearly discriminate this model from 
 standard cosmology with cosmological constant.     
\end{abstract}

\pacs{95.36.+x, 98.80.-k, 98.80.Es}  

\maketitle


Recent cosmological observations \cite{SN1,SN2,WMAP,SDSS} indicate that the universe is
undergoing a phase of {\it accelerated} expansion. The fact that
the present rate of expansion is accelerating rather than decelerating
poses one of the most important problems of modern cosmology. 
Indeed, 
in  Standard  Cosmology our universe starts expanding after the Big Bang,
 but the attractive nature of gravity for ordinary matter and radiation 
necessarily slows down the expansion rate. In order to have
acceleration, Einstein's equations
require the universe to be dominated  by some
sort of non-ordinary energy (usually called {\it dark energy})
with the particular property of having {\it negative} pressure.

Although its nature  is unknown, a simple phenomenological 
description in which dark energy is understood as a 
cosmological constant,  i.e. a perfect fluid 
with 
equation of state, $p_{\Lambda}=-\rho_{\Lambda}$, 
where $\rho_{\Lambda}$ and $p_{\Lambda}$ are 
the energy density and pressure respectively, seems to fit observations 
with very good precision ($\Lambda$CDM model). Thus,  $\Lambda$CDM suggests 
that around a 70$\%$ of the energy density of the
universe today would be in the form of dark energy, whereas the remaining
30$\%$ would be non-relativistic matter (the contribution from radiation and
curvature being negligible) \cite{WMAP}. 

However the fact that today matter and 
dark energy have comparable contributions to the energy density  (both around
($10^{-3}$ eV)$^4$ in $\hbar=c=1$ units) turns out to be difficult to understand
if dark energy is a true cosmological constant. Indeed, the energy density of a
cosmological constant remains  constant throughout the history of the universe,
whereas those of the rest of components (matter or radiation) grow as we go back in time.
Then the question arises as to whether it is a {\it coincidence} (or not) that 
they have comparable values today when they have differed 
by many orders of magnitude in  the past.  
In addition, the cosmological constant
exhibits another related problem. Its scale (around $10^{-3}$ eV) 
is more than 30 orders of magnitude smaller than the scale of the 
other dimensional constant appearing in the gravitational equations, 
$G=M_P^{-2}$ with $M_P\sim 10^{19}$ GeV, and
it is also difficult to explain from particle or other known physics.

In order to avoid these problems several models have been proposed
in which dark energy is a dynamical component rather than a 
cosmological constant.
Such models are usually based on cosmological scalar fields or 
 modifications of Einstein's gravity 
\cite{quintessence1,k-essence,f(R),DGP}. However, 
in order to have 
acceleration at the right time, they typically introduce unnatural 
dimensional scales, resulting once again in fine tuning or 
coincidence problems.

In this paper we consider a completely different  type of dark energy model
which is not based on scalar fields, but in the dynamics of a vector field.
Unlike previous works \cite{Kiselev,Boehmer}, 
it is  shown that vector fields can 
give rise to periods of acceleration even in the absence of potential terms
(see \cite{Ferreira} for a general analysis). 
The existence of such solutions 
does not rely on the introduction of complicated functions of  the 
fields and its derivatives, but can be obtained with the simplest kinetic
terms, including two fields and two derivatives, so that the model does 
not contain
dimensional parameters. Furthermore, we show that the required
initial conditions for the vector fields are natural.

Let us start by writing the action of our 
vector-tensor theory of gravity containing only 
two fields and two derivatives and without potential terms \cite{Will}:
\begin{eqnarray}
S&=&\int d^4x \sqrt{-g}\left(-\frac{R}{16\pi G}\right.\nonumber\\
&-&\left.\frac{1}{2}\nabla_\mu A_\nu\nabla^\mu A^\nu
+\frac{1}{2} R_{\mu\nu}A^\mu A^\nu\right).
\label{action}
\end{eqnarray}
Notice that the theory contains no free parameters, the only dimensional
scale being  Newton's constant.   
The numerical factor
in front of the vector kinetic terms can be fixed 
by the field normalization.
Also notice that  
 $R_{\mu\nu}A^\mu A^\nu$ can be written as a combination of derivative terms as 
$\nabla_\mu A^\mu\nabla_\nu A^\nu-\nabla_\mu A^\nu\nabla_\nu A^\mu$.

The classical equations of motion derived from the action in (\ref{action})
are the Einstein's and
 vector field equations:
\begin{eqnarray}
R_{\mu\nu}-\frac{1}{2}R g_{\mu\nu}&=&8\pi G (T_{\mu\nu}+T_{\mu\nu}^A) \label{eqE}\\
\Box A_\mu + R_{\mu\nu}A^\nu&=&0, \label{eqA}
\end{eqnarray}
where $T_{\mu\nu}$ is the conserved energy-momentum tensor for matter and radiation and
$T_{\mu\nu}^A$ is the energy-momentum tensor coming from the vector field.
In this work we shall solve these equations for the simplest isotropic and
 homogeneous flat cosmologies. 
Thus, we assume that 
the spatial components of the vector field vanish, so that 
$A_\mu=(A_0(t),0,0,0)$ and, therefore, the 
space-time geometry will 
be given by the flat Robertson-Walker metric:
\begin{equation}
ds^2=dt^2-a^2(t)\delta_{ij}dx^idx^j,
\label{metric}
\end{equation}
For this metric (\ref{eqA}) reads:
\begin{eqnarray}
\ddot{A}_0+3H\dot{A}_0-3\left[2H^2+\dot{H}\right]A_0=0,\label{fieldeq0}
\end{eqnarray}
where $H=\dot a/a$ is the Hubble parameter. 

Assuming that the universe has gone through  radiation and 
matter phases in which
the contribution from dark energy was negligible, we can easily solve
these equations in those periods just taking $H=p/t$, with 
 $p=1/2$ for radiation and  $p=2/3$ 
for matter  eras respectively, which is
equivalent to assuming that $a\propto t^p$.  In that case, the
above equation has a growing and a decaying solution:
\begin{eqnarray}
A_0(t)=A_0^+t^{\alpha_+}+A_0^-t^{\alpha_-},\label{fieldsol}
\end{eqnarray}
with $A_{0}^\pm$ constants of integration and $\alpha_{\pm}=-(1\pm 1)/4$
 in the radiation era, and
 $\alpha_{\pm}=(-3\pm\sqrt{33})/6$ in the matter era.

On the other hand, the $(00)$ component of Einstein's equations reads:
\begin{eqnarray}
H^2=\frac{8\pi G}{3}
\left[\sum_\alpha \rho_\alpha+\rho_{A}\right]
\label{Friedmann}\end{eqnarray}
with $\alpha=M,R$ and:
\begin{eqnarray}
\rho_{A}&=&\frac{3}{2}H^2A_0^2+3HA_0\dot A_0-\frac{1}{2}\dot A_0^2.
\end{eqnarray}

Using the
growing mode solution in (\ref{fieldsol}), we obtain:
\begin{eqnarray}
\rho_{A}= \rho_{A0} a^\kappa ,
\end{eqnarray}
with $\kappa=-4$ in the radiation era and 
$\kappa=(\sqrt{33}-9)/2 \simeq -1.63$ in 
the matter era. Thus the energy density of the vector field 
starts scaling as radiation
 at early times, so that $\rho_A/\rho_R=$ const.  
However when the universe enters its matter era, $\rho_A$ starts growing 
relative to 
$\rho_M$ eventually overcoming it at some point, in which the dark energy 
vector  
field would become
the dominant component. From that point on,  we cannot obtain analytic
solutions to the field equations. In Fig. 1 we show the numerical solution
to the exact equations, which confirms our analytical estimates in the 
radiation and matter eras.  Notice that  
since $A_0$ is essentially constant during 
the radiation era, solutions do not depend on the precise initial time
at which we specify it.  Thus, 
once the present value of the Hubble parameter $H_0$
and the constant $A_0$ during radiation (which fixes the total matter
density $\Omega_M$) are specified, 
the model is completely 
determined. In other words, this model contains the same
number of parameters as $\Lambda$CDM, i.e. the minimum number of parameters
of a cosmological model with dark energy. As seen from
Fig.1  the evolution of the universe ends at a finite time $t_{end}$ with a 
 singularity in which
$a\rightarrow a_{end}$ with $a_{end}$ finite, $\rho_{DE}\rightarrow \infty$,  
$p_{DE}\rightarrow -\infty$ and $A_0(t_{end})=M_P/(4\sqrt{\pi})$. This 
corresponds to a Type III singularity according to the classification
in \cite{Nojiri}.

\begin{figure}[h]
\vspace{0.3cm}
\begin{center}{\epsfxsize=8.0 cm\epsfbox{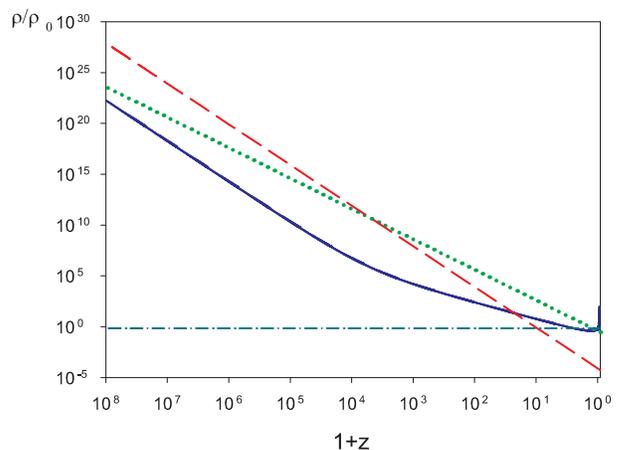}}
\caption{\small Evolution of energy densities for the best fit model. 
Dashed (red) for
radiation, dotted (green) for matter and solid (blue) for vector 
dark energy. We show also for comparison the cosmological constant
density in dashed-dotted line. We see the rapid growth of the vector 
dark energy
contribution at late times approaching the final singularity.}
\end{center}
\end{figure}

We can also calculate the effective equation of state for dark energy
as:
\begin{eqnarray}
w_{DE}=\frac{p_A}{\rho_A}=\frac{-3\left(\frac{5}{2}H^2+\frac{4}{3}\dot{H}\right)A_0^2+HA_0\dot{A}_0
-\frac{3}{2}\dot{A}_0^2}{\frac{3}{2}H^2A_0^2+3HA_0\dot A_0-\frac{1}{2}\dot A_0^2}
\end{eqnarray}
Again, using the approximate solutions in (\ref{fieldsol}), we obtain;
\begin{eqnarray}
w_{DE}=\left \{
\begin{array}{l}
\frac{1}{3}\;\;\;\;\;\;\;\;\;\;\;\;\;\;\;\;\;\;\;\;\;\;\;\;\;\;\;\;\;\; \mbox{radiation era}\\
\frac{3\sqrt{33}-13}{\sqrt{33}-15}\simeq -0.457\;\;\;\;\mbox{matter era}
\end{array}
\right.
\end{eqnarray}
 After
dark energy starts dominating, the equation of state abruptly falls towards 
$w_{DE}\rightarrow -\infty$ as the universe approaches $t_{end}$. 
As shown in Fig. 2 the equation of state can cross the so called phantom 
divide, so that we can have $w_{DE}(z=0)<-1$.
In Fig. 3, we show the evolution of the $A_0$ component. 


\begin{figure}[h]
\vspace{-0.2cm}
\begin{center}{\epsfxsize=7.1 cm\epsfbox{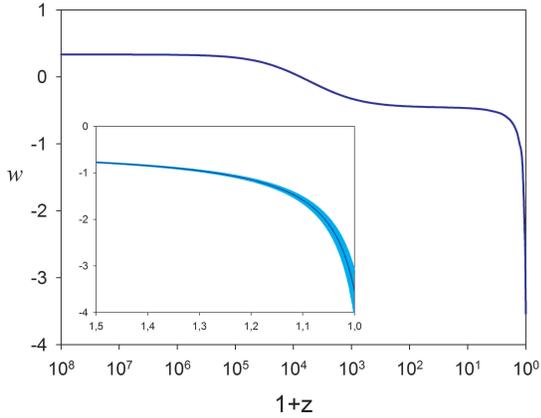}}
\caption{\small Evolution of dark energy equation of state for
the best fit model. The lower panel shows the 1$\sigma$ confidence interval.}
\end{center}
\end{figure}

\begin{figure}[h]
\begin{center}{\epsfxsize=7.3 cm\epsfbox{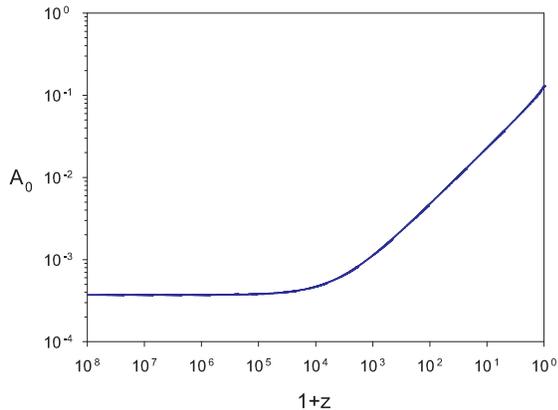}}
\caption{\small Evolution of $A_0$ in $M_P$ units for
the best fit model.}
\end{center}
\end{figure}

In order to confront the  predictions of the model with observations
of high-redshift type Ia supernovae, 
we have calculated the distance modulus as a function of redshift.
Comparing $\mu_{th}(z)$ with its observational
 value in a given data set will enable us to carry out a 
$\chi^2$ statistical analysis.
 For this purpose, we have considered two sets of supernovae: 
the Gold set \cite{Gold}, 
containing 157 points with $z < 1.7$, and  
the more recent SNLS  data set \cite{SNLS}, comprising 115 
supernovae but with lower redshifts ($z < 1$).

In Table 1 we show the results for the best fit together 
with its corresponding $1\sigma$  intervals for the 
two data sets. We also
show for comparison the results for a standard
$\Lambda$CDM model. We see that the vector model (VCDM) fits the data
considerably
better than  $\Lambda$CDM (at more than $2\sigma$)  in the Gold 
set, whereas the situation is reversed in the SNLS set. This is just 
a reflection of the well-known $2\sigma$ tension \cite{tension}
between the two 
data sets. The best fit parameters for the
VCDM model are identical for the two data sets with small differences in
the confidence intervals. Compared with $\Lambda$CDM, we see that VCDM
favors a younger universe (in $H_0^{-1}$ units)
with larger matter density. In addition, 
the deceleration-aceleration transition takes place at a lower redshift
in the VCDM case. Another important difference arises in the present value
of the equation of state  with $w_0=-3.53^{+0.46}_{-0.57}$ 
which clearly excludes the cosmological constant value $-1$. Future
surveys \cite{Trotta} are expected to be able to measure $w_0$ at the 
few percent level
and therefore could discriminate between the two models. 


We have also compared with other parametrizations
for the dark energy equation of state. Thus for instance, 
taking $w_{DE}(z)=w_0+w_1 z(1+z)^{-1}$  \cite{CPL1}, 
we find $\chi^2=173.5$ for the Gold set. Since this is a 
three-parameter fit, in order to compare with the one-parameter fits 
of VCDM or  $\Lambda$CDM, we use the reduced chi-squared: 
$\chi^2/d.o.f=1.108$ for
VCDM, $\chi^2/d.o.f=1.127$ for the $(w_0,w_1)$ parametrization and 
$\chi^2/d.o.f=1.135$ for $\Lambda$CDM. As a matter of fact, to  our 
knowledge best, VCDM provides 
the best fit to date for the Gold data set, since the  
  oscillatory four-parameter model previously  reported in \cite{models} 
still has $\chi^2/d.o.f=1.115$.

The evolution of dark energy for the best-fit model is plotted in Figs. 1-3. 
We see that unlike the 
cosmological constant case, throughout the radiation era
 $\rho_{DE}/\rho_R\sim 10^{-6}$. Notice that although the 
onset of cosmic acceleration depends on the value of $A_0$ during that era,  
for the best-fit $A_0=3.71 \times 10^{-4}$ $M_P$, which  is 
relatively close to the Planck scale and could arise naturally  
in the early universe without the need of introducing extremely small 
parameters.
\begin{table}
\begin{center}
\begin{tabular}{|c|c|c|c|c|}
\hline
  & & & &\\ 
 & VCDM  & $\Lambda$CDM  & VCDM & $\Lambda$CDM \\
& \footnotesize{Gold}  &\footnotesize{Gold} & 
\footnotesize{SNLS}   & \footnotesize{SNLS} \\
& & & &\\
\hline & & & & \\$\Omega_M$ &$0.388^{+0.023}_{-0.024}$ & $0.309^{+0.039}_{-0.037}$ & 
$0.388^{+0.022}_{-0.020}$ & $0.263^{+0.038}_{-0.036}$
\\& & && \\ 
\hline & & & &\\ $w_0$ &$-3.53^{+0.46}_{-0.57}$& $-1$ & $-3.53^{+0.44}_{-0.48}$ & $-1$
\\& & & & \\
\hline  & & & &\\$A_0$ & $3.71^{+0.022}_{-0.026}$ &--- & $3.71^{+0.020}_{-0.024}$&---
\\\footnotesize{$(10^{-4}\;M_P)$}& & & & \\
\hline  & & & &\\$z_T$ &$0.265^{+0.011}_{-0.012}$ &$0.648^{+0.101}_{-0.095}$ & $0.265^{+0.010}_{-0.012}$&$0.776^{+0.120}_{-0.108}$ 
\\& & & & \\
\hline  & & & &\\$t_0$   &$0.926^{+0.026}_{-0.023}$ & $0.956^{+0.035}_{-0.032}$& 
$0.926^{+0.022}_{-0.022}$ &$1.000^{+0.041}_{-0.037}$ 
\\\footnotesize{$(H_0^{-1})$}& &  & & \\
\hline & & & &\\$t_{end}$ &$0.976^{+0.018}_{-0.014}$ & --- &$0.976^{+0.015}_{-0.013}$ & ---
 \\\footnotesize{$(H_0^{-1})$}& & & &\\
\hline & & & &\\ $\chi^2_{min}$ & 172.9& 177.1 &115.8 & 111.0
\\ & & & &\\ 
\hline
\end{tabular}
\caption{\footnotesize{Best fit parameters with $1\sigma$ 
intervals for the vector model (VCDM) and
the cosmological constant model ($\Lambda$CDM) for the Gold (157 SNe)
and SNLS (115 SNe) data sets. $w_0$
denotes the present equation of state of dark energy. $A_0$ is 
the constant value of the vector field component during radiation. 
$z_T$ is the deceleration-aceleration transition redshift.
$t_0$ is the age of the universe in units of the present
Hubble time. $t_{end}$ is the duration of the universe in 
the same units.}}
\end{center}
\end{table}

When comparing the parameters obtained
from SN Ia (Table 1) with predictions coming 
from  CMB anisotropies or baryon acoustic
oscillations \cite{BAO}, 
it is important to keep 
in mind that  such predictions 
are obtained  after
a data process which involves the use of a particular model 
for dark energy, which in most  cases is $\Lambda$CDM, as a fiducial
model.  This is a good approximation for models with (nearly)
constant equation of state \cite{BAO}, but could not be 
 a priori justified   in our case since
$w_{DE}(z)$ has a strong redshift 
dependence \cite{Dick}. 

In this work we have only considered the time component of
the vector field. The presence of  spatial
components  could, in principle,  have adverse effects.  
However, we have found that 
the energy density
of the spatial part decays as $a^{-8}$ during radiation and matter eras, 
i.e. much faster than the temporal contribution, so that 
it will not dominate at late times. On the other hand,
we have calculated the evolution of $(p_\parallel-p_\perp)/\rho$, where
  $p_\parallel$ is the pressure along the direction of the spatial
component and $p_\perp$ is the transverse pressure, and we have found 
that this quantity decays very fast during the matter and radiation eras. 
Accordingly, we do not 
expect the generation of large anisotropies.

So far we have only considered the homogeneous model. In 
order to study the model stability we have considered the evolution 
of metric and vector field perturbations. Thus, we obtain the dispersion
relation and  the propagation speed of  scalar, vector and tensor modes.  
For all of them we obtain $v=(1-16\pi G A_0^2)^{-1/2}$ which is
 real throughout the universe evolution, since the value $A_0^2=(16\pi G)^{-1}$
exactly corresponds  to that at the  final singularity. Therefore
 the model does not exhibit exponential instabilities. As shown in
\cite{Mukhanov}, the fact that the propagation speed is faster than $c$ 
does not necessarily imply inconsistencies with causality. We have 
also considered the evolution of scalar perturbations in the vector field 
generated by scalar metric perturbations during the matter and radiation eras, 
and found that the energy density contrast
$\delta \rho_A/\rho_A$ is constant on super-Hubble
scales, whereas it oscillates with growing amplitude as $a^2$ 
in the radiation  era and as $\sim a^{0.3}$ in the matter era for 
sub-Hubble scales.  
Therefore  again, we do not find exponentially 
growing modes.

The model proposed in this work can be considered as an effective 
description of dark energy on cosmological scales. Extending the 
applicability range to smaller scales 
requires consistency with  local gravity tests. Indeed, we can see that
 for the model in (\ref{action}), the static
post-Newtonian parameters agree with those of General Relativity \cite{Will}, i.e. 
$\gamma=\beta=1$. 
For the parameters  
associated to preferred frame effects we get: $\alpha_1=0$ and
$\alpha_2=8\pi  A_\odot^2/M_P^2$ where $A_\odot^2$ is the norm of the 
vector field at the solar system scale. Current limits 
$\alpha_2 \;\lsim 10^{-4}$ (or $\alpha_2\;\lsim 10^{-7}$ for static vector 
fields during
solar system formation) then impose a bound $A_\odot^2\lsim \, 
10^{-5}(10^{-8})
\,M_P^2$, which could conflict with the model predictions, since
the present (Solar system formation)  values on cosmological scales are: 
$1.3\times 10^{-1}\,M_P$ ($7.5\times 10^{-2}\,M_P$).
However, notice that the cosmological values 
do not need to agree with those at lower scales. The latter 
will be determined by the mechanism
that generated this field in the early universe characterized by its 
primordial spectrum of perturbations, and the
subsequent evolution in the formation of the galaxy and Solar system. 
Concerning the potential presence of quantum instabilities in the model, 
 in \cite{Gripaios} the condition in order to ensure positive norm
Hilbert space are obtained.  With our Riemann tensor 
sign convention, we see that such a condition is indeed satisfied in 
the model (\ref{action}).

 In conclusion, the results of this
work show that vector theories offer an accurate 
phenomenological description of dark energy in which fine tuning problems
 could be  avoided.

{\em Acknowledgments:} This work has been  supported by
DGICYT (Spain) project numbers FPA 2004-02602 and FPA
2005-02327, UCM-Santander PR34/07-15875 and by CAM/UCM 910309. 
J.B. aknowledges support from MEC grant
BES-2006-12059.


\vspace{-0.5cm}
\begin{thebibliography}{99}
\vspace{-0.4cm}
\bibitem{SN1} S. Perlmutter et al., {\it Astrophys. J.} {\bf 517}, 565 (1999)
\bibitem{SN2} A.G. Riess et al., {\it Astron. J.} {\bf 116}, 1009 (1998) and
{\bf 117}, 707 (1999)
\bibitem{WMAP} D. N. Spergel et al. \emph{Astrophys. J. Suppl.} \textbf{148},
175, (2003) and  astro-ph/0603449
\bibitem{SDSS} M. Tegmark et al., {\it Phys. Rev.} {\bf D69}: 103501, (2004).
\bibitem{quintessence1} C. Wetterich, {\it Nucl. Phys.} {\bf B302}, 668 (1988); 
R.R. Caldwell, R. Dave and P.J. Steinhardt, {\it Phys. Rev. Lett.}
 {\bf 80}, 1582 (1998)
\bibitem{k-essence} C. Armendariz-Picon, T. Damour and V. Mukhanov, 
{\it Phys. Lett.} {\bf B458}, 209 (1999)
\bibitem{f(R)} S.M. Carroll, V. Duvvuri, M. Trodden, M.S. Turner, {\it 
Phys. Rev.} {\bf D70}: 043528, (2004) 
\bibitem{DGP} G. Dvali, G. Gabadadze and M. Porrati, 
{\it Phys. Lett.} {\bf B485}, 208 (2000)
\bibitem{Kiselev} V.V. Kiselev, {\it Class. Quant. Grav.} 
{\bf 21}: 3323, (2004). 
\bibitem{Boehmer} C. Armendariz-Picon, {\it JCAP} {\bf 0407}: 007, (2004);
C.G. Boehmer and T. Harko, {\it Eur. Phys. J.} 
{\bf C50}: 423, (2007); M. Novello, et al. {\it Phys. Rev.} {\bf D69}:
127301, (2004); T. Koivisto and D.F. Mota, e-Print: arXiv:0707.0279;
\bibitem{Ferreira} P.~G.~Ferreira, et al., {\it Phys. Rev.} {\bf D75}: 044014 (2007) 
\bibitem{Will} C. Will, {\it Theory and experiment in gravitational physics},
Cambridge University Press, (1993)
\bibitem{Nojiri} S.~Nojiri, S.~D.~Odintsov and S.~Tsujikawa,
  {\it Phys.\ Rev.} {\bf D71} (2005) 063004
\bibitem{Gold} A.G. Riess at al. {\it Astrophys.J.} {\bf 607}, 665
(2004)
\bibitem{SNLS} P. Astier et al.,  {\it Astron. Astrophys.}
{\bf 447}: 31-48, (2006). 
\bibitem{tension} S. Nesseris and L. Perivolaropoulos, {\it JCAP} 
{\bf 0702}: 025, (2007). 
\bibitem{Trotta} R. Trotta and R. Bower, {\it Astron. Geophys.}
{\bf 47}: 4:20-4:27, (2006)
\bibitem{CPL1} M. Chevalier and D. Polarski, {\it Int. J. Mod. Phys.} 
{\bf D10}, 213 (2001); E.V. Linder, {\it Phys. Rev. Lett.} {\bf 90} 091301 (2003) 
\bibitem{models}R. Lazkoz, S. Nesseris, L. Perivolaropoulos, {\it JCAP} 
{\bf 0511}:010, (2005) 
\bibitem{BAO} D.J. Eisenstein et al. {\it Astrophys. J.} {\bf 633}: 
560-574, (2005) 
\bibitem{Dick} J. Dick, L. Knox and M. Chu, {\it JCAP} {\bf 0607}: 
001, (2006). 
\bibitem{Mukhanov}
  E.~Babichev, V.~Mukhanov and A.~Vikman, JHEP {\bf 0802}, 101 (2008)
\bibitem{Gripaios}
  B.~M.~Gripaios,
  {\it JHEP} {\bf 0410} 069 (2004) 
\end{thebibliography}
\end{document}